# Site Selection for the Second Flyeye Telescope: A Simulation Study for Optimizing Near-Earth Object Discovery

D. Föhring[a*], L. Conversi[a], M. Micheli[a], E. Dölling[b], P. Ramirez Moreta[c]

[a] *ESA PDO NEO Coordination Centre (Via Galileo Galilei, 1, 00044 Frascati (RM), Italy)*

[b] *ESA/ESOC (Robert-Bosch-Str. 5 64293 Darmstadt, Germany)*

[ci] *ESA ESAC (Camino Bajo del Castillo s/n, 28692 Villafranca del Castillo, Madrid, Spain*

## Abstract

The European Space Agency (ESA) is developing a network of wide-field survey telescopes, named Flyeye, to improve the discovery of Near-Earth Objects (NEOs). The first telescope in the network will be located in the Northern Hemisphere on Mount Mufara (Italy), and a second Flyeye telescope, featuring increased detection capabilities, has just started the critical design phase.

The potential location for the second Flyeye telescope is investigated by performing simulations of NEOs on impacting trajectories. Approximately 3000 impacting asteroids of two absolute magnitudes (H=25 and H=28) were propagated and tested for detectability by major existing surveys (Catalina, Pan-STARRS, ATLAS), the upcoming Vera Rubin Observatory (LSST), and possible Flyeye locations. Chile, South Africa, and a second facility in the Northern Hemisphere were considered. For each observatory, their past or planned pointing strategies were taken into account in the simulation.

Before LSST deployment, a single Flyeye in the Southern Hemisphere performs similarly to a telescope in the Northern Hemisphere. When combined, having one telescope in the north and one in the south maximizes detections and number of unique objects detected. After LSST, southern and northern Flyeye telescopes remain complementary. Overall, simulations show that a second Flyeye in the south complements a Flyeye telescope in the north both before and after LSST. A Flyeye located at La Silla would take advantage of the excellent atmospheric conditions, while allowing a balance of assets across hemispheres.

## 1. Introduction

The building of ESA's network of survey telescopes dedicated to discovering smaller asteroids on a direct collision course with the Earth is currently on its way. The telescopes will be called "Flyeye", due to their unique optical design which splits the light into multiple cameras resembling the fly's eye (Cibin et al., 2019; Marchiori et al., 2022). The first unit in the network, Flyeye-1, will be a 1-metre class telescope with 16 cameras and a 6.7°×6.7° field of view, capable of performing a complete scan of the observable sky down to V~21.5 every 2-3 nights. Currently, there is funding to build other large field of view telescopes (e.g. Gregori et al., 2023) for ESA's network. This network will be of particular importance for Europe and ESA, because currently most NEO surveys are in the United States, funded by NASA. The three surveys that currently contribute the most to new NEO discoveries are: Pan-STARRS in Hawaii, USA (Kaiser et al., 2002); Catalina Sky Survey in Arizona, USA (Larson et al., 1998); and ATLAS, which has telescopes in the USA, South Africa and Chile (Tonry et al., 1998) and another currently under construction in Spain. These survey telescopes automatically scan the sky every night looking for new asteroids. A major change is expected to happen to the field once the Vera Rubin Observatory Legacy Survey of Space and Time (LSST) will become operational. This telescope will be located in Chile and will scan the sky in 6-bands ($u, g, r, i, z,$ and $y$) down to a predicted depth of ~24.7 mag in $r$. The telescope will perform ~1000 visits per night covering at least 18 000 square degrees (Ivezić et al. 2019; Jones et al., 2015). It is expected to see first light in 2025.

---

[*] Corresponding author.
*Email address:* `dora.fohring@ext.esa.int`

Given the context above, we wanted to answer the following questions:

- Will LSST detect the majority of imminent impactors in the Southern Hemisphere? Is it worth building more survey telescopes in the South?
- If the majority of survey telescopes are in the Northern Hemisphere, is it worth building more survey telescopes in the North?
- Does longitude affect detection rate?

To answer these questions, we performed simulations as described in the following section. Previous simulations on the detection rate of the Flyeye-1 telescope have been done before (Ramirez Torralba et al., 2019); however, this work is the first to consider the impact of the actual observing strategies and the effect of contemporaneous observations by other observatories, while focussing on the "population" of impacting NEOs.

## 2. Method

A population of ~3000 impacting asteroids, provided by S. Chesley (Chesley et al., 2019) were simulated. These were all assumed to be impacting over the course of a single year. Two regimes were considered: one where the impactors had an H-magnitude = 25, and another where H = 28. Assuming a typical albedo of 0.15, this corresponds to asteroid sizes of 35 m and 8.7 m, respectively. We simulated which of these impactors would be detected in one year by each of the current surveys. Next, we simulated which of these impactors would be detected by LSST and the Flyeye-1. Finally, we considered the Flyeye-2 in the following locations:

- A: In the Northern Hemisphere, co-located with Flyeye-1 in Mufara, Italy (abbreviated Muf)
- B: In the Southern Hemisphere, in La Silla, Chile, at approximately the same longitude as LSST (observatory code 809)
- C: In the Southern Hemisphere, in Sutherland, South Africa, in order to have a ~6 h longitudinal separation from LSST (observatory code B31)

We sought to determine from which of these locations the Flyeye-2 would find the most impactors, and the greatest number of unique impactors. We also analysed what effect of LSST would have on these observations. This analysis was used to provide a recommendation for a potential location of the Flyeye-2.

### 2.1 Observing Strategies

For existing surveys, their past pointings were obtained from the Minor Planet Center (MPC), for a single year, chosen to be 2020. The two newer ATLAS telescopes in Chile and South Africa are an exception, as their pointings are not yet available from the MPC and have only been running since 2022. For these two, the pointings were obtained from the ATLAS portal[1]. For LSST, one year of simulated pointings were taken from their most recent simulated survey strategy at the time of writing, which is publicly available, called *draft_connected_v2.99_10yrs*[2]. For the Flyeye-1, the current scheduler can only take into account observations from previous nights if they have been reported to the MPC, so for the purpose of the investigation we had to simulate a strategy for it. For the Flyeye-1, LSST and the two newer ATLAS telescopes, where individual pointings were used, an asteroid was flagged as discovered only after it was detected at least four times. A detection was defined as when object was in the observed field of view with a magnitude less than the limiting magnitude of the telescope. The detections were counted per object, not per night, which is particularly important for LSST, whose strategy consists of revisiting pairs of fields every 3 – 4 nights.

*2.1.1 Flyeye Strategy*

The observing strategy adopted for the Flyeye telescopes was created as follows:

The entire celestial sphere was divided into a tessellation grid of equal regions corresponding to the Flyeye field of view. From this grid, points that were less than 15 degrees from the Galactic Plane were omitted. The distance was fine-tuned to produce a region of similar width as the region with the lower number of visits for LSST, as can be seen in e.g. Figure 3 of Chesley and Veres (2017). Observation Strategy Definition (OSD) files were then produced for each night from the tessellation grid using Python's Astropy module (The Astropy Collaboration et al., 2022) according to the following method: for each Flyeye location, the latitude, longitude and elevation of the observatory were as defined in Table 1.

| Location | Longitude (°) | Latitude (°) | Elevation (m) |
|---|---|---|---|
| **Mufara** | 14.0201 | 37.8689 | 1865 |
| **La Silla (809)** | 289.251 | -30.244 | 2400 |
| **Sutherland (B31)** | 20.811 | -32.202 | 1798 |

Table 1. Longitude, latitude and elevation of the sites used in the simulations

A night was defined for each day of the year for each location, between astronomical twilights, when the Sun is -18° or lower below the horizon. To simulate the effect of closures due to technical maintenance, the nights were uniformly shortened by 10% before sunrise. Using a cadence of 80 s (60 s exposure time plus overheads), we calculated how many observations can fit into a single night. Every data point in the tessellation grid was repeated in segments of 15 elements, corresponding to a 20-minute revisit time, 4 times. This list of pointings was looped over repeatedly, starting at the beginning of the first night of the first day, and checked whether the field was visible. If it was, it was included in the OSD file for the night. The time was then incremented by the cadence, and the observability of the next field was checked. This was repeated for the duration of the night. The limitation of this method is that, especially towards the end of the night, it may become impossible to get the 4$^{th}$ observation scheduled for the night. In real life, the Flyeye scheduler will have knowledge of the successful observations it carried out the nights before and will be able to adjust the schedule accordingly. Simulating this was beyond the scope of this work, but a simple approximation was performed: upon successfully scheduling the 4$^{th}$ observation, the pointing was saved in a list of pointings of that night. In the subsequent night, these points were excluded from the scheduling. Figure 1 shows examples of typical simulated night-to-night pointings for the Flyeye for the Northern hemisphere. Figure 2 shows the overall area covered in the sky over a year from the Northern and Southern Hemispheres. The moon phase was not taken into account. For the survey strategy consisting of 2 Flyeye telescopes in the North, a similar strategy was generated but with a cadence of 40 s, to emulate the doubling of the observing capability. The limiting magnitude of V=21.5 was kept the same.

---

[1] https://astroportal.ifa.hawaii.edu/atlas/pointing/
[2] https://s3df.slac.stanford.edu/data/rubin/sim-data/sims_featureScheduler_runs3.0/draft2

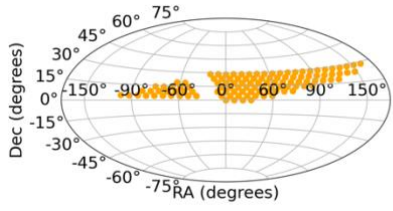
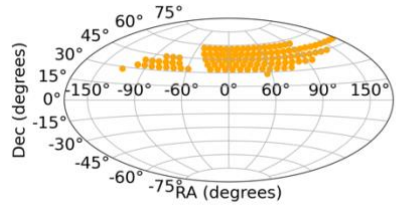
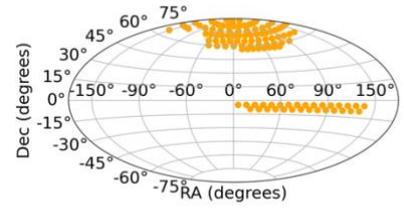

Fig. 1 (a) Night 1    Fig. 1 (b) Night 2    Fig. 1 (c) Night 3

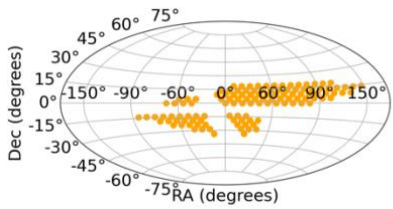
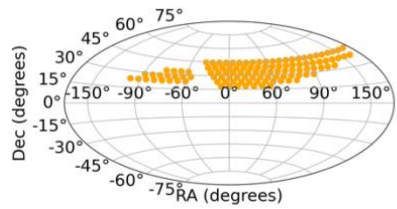
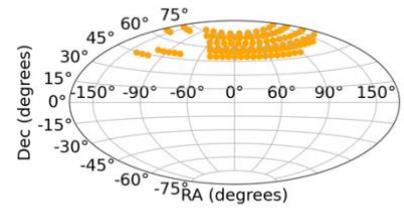

Fig. 1 (d) Night 4    Fig. 1 (e) Night 5    Fig. 1 (f) Night 6

Figure 1 (a) -- (f): sequence of simulated consecutive nights for the Flyeye in the Northern Hemisphere.

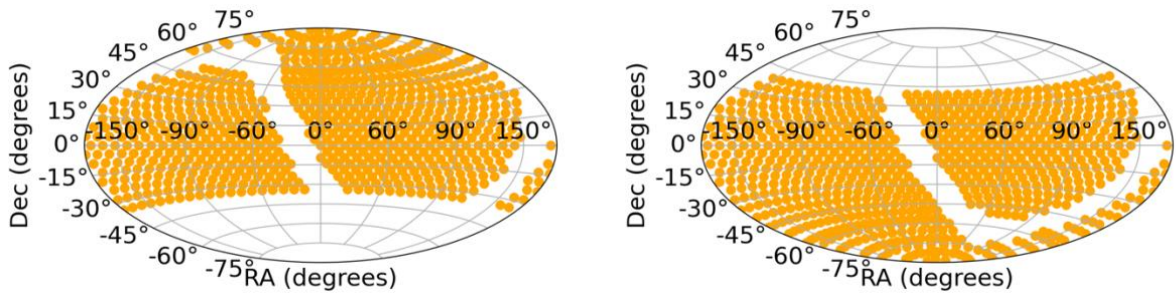

Figure 2. Yearly coverage map of the Flyeye in the North (left) and South (right). Over the course of the year, the Flyeye will cover the area shown in orange uniformly.

*2.1.2 LSST Strategy*

The LSST strategy consists of a pair of exposures separated by 20 minutes, followed by another pair, 3 - 4 days later. Because of this separation between pairs, a successful detection depends on the ability for the LSST pipeline to perform the linkage: in this work, it is assumed that the algorithm used by LSST has a 100% efficiency in performing the linkage.

*2.2 Orbital Propagation*

Orbital propagation consisted of updating the positions of the impactors in the sky from their orbital elements using the ephemeris computation service provided by JPL Horizons[3] (Giorgini et al., 1996), accessed via their API. Orbital propagation was carried out in intervals of 1 day starting 60 days before impact, through consecutive queries for each impactor for three locations, Mufara, La Silla (809) and Sutherland (B31). These queries were saved and subsequent analyses for the same location were based on reading the relevant files. The 1-day interval was chosen for several reasons. Over the 60-day time span analysed, the median apparent velocity of an impactor was found to remain below 1 "/min, only increasing sharply within 24 hours before impact. At this speed, the impactor would remain in the field of view of a survey telescope throughout the night. This means that as long as a telescope pointed to that region at any time during the night, it would have the impactor in its field of view. This simplified calculations and optimised computing time. Additionally, the pointing files available from the MPC only recorded positions with an accuracy of one day, so greater temporal precision could not have been used effectively.

## 2.3 Asteroid Detection

### 2.3.1 Limiting Magnitudes

During orbital propagation, the apparent magnitude of each object was checked against the limiting magnitude of the telescope. This was done in two passes. Each telescope was given a maximum limiting magnitude to speed up calculations. If the object, on a given night, was fainter than this value, it was passed over. Otherwise, it was taken to the next step, where it was compared to the position of the field and its actual limiting magnitude. For telescopes with pointings from the MPC, their limiting magnitudes were taken as reported to the MPC. For the Flyeye the limiting magnitude was assumed to be always 21.5. For LSST, the simulated limiting magnitudes were given in filters, and this was corrected to *V* using offsets as specified in Table 2. The correction values used are the bandpass corrections from ESA's orbital dynamics system, Aegis (Fenucci et al., 2024), and correspond to the colour offsets most commonly used in all orbital dynamics centres (see e.g. Williams, 2013).

| Filter | u' | g' | r' | i' | z' | y' |
|---|---|---|---|---|---|---|
| Offset | +2.50 | -0.28 | +0.23 | +0.39 | +0.37 | +0.36 |

Table 2. Offsets in magnitudes of LSST filters relative to the standard V filter.

For the two newer ATLAS telescopes, instead of the limiting magnitude, the Zero Point magnitude was given for each observation. To convert this to limiting magnitude we assumed that the maximum possible limiting magnitude for ATLAS is 22, where the telescope was operating at maximum sensitivity, and the limiting magnitude at maximum sensitivity was assumed to be 19.6. When the reported limiting magnitude was less than 22, it meant that the telescope was operating at less than maximum sensitivity, and we calculated the delta between 22 and the reported Zero Point magnitude. This delta was then subtracted from the assumed maximum limiting magnitude of 19.6 to give an estimate for the limiting magnitude for the specific observation.

---

[3] https://ssd.jpl.nasa.gov/horizons/

*2.3.2 Trailing*

As the impactor approaches Earth, its apparent velocity in the sky will increase, causing the Point Spread Function (PSF) to appear trailed in the typical survey exposure times. This trailing will cause the total flux from the object to be spread out over more pixels along the detector, resulting in an overall dimming of the object compared to its expected brightness. Additionally, an extra detection loss is introduced depending on how well the algorithm can identify sources in trailed images, for example when a PSF-like filter is used on a trailed source.

Taking *M* to be the original magnitude of the object, the trailed magnitude, $M_{trail}$, can be expressed as:

$M_{trail} = M + DM_{trail}$,

where $DM_{trail} = DM_{detect} + DM_{SNR}$,

wiith $DM_{trail}$ being the total trailing loss, which consists of $DM_{detect}$, the detection loss, and $DM_{SNR}$, the magnitude loss due to the SNR trailing.

The detection loss is described by the function (Ivezić et al., 2019):

$DM_{detect} = 1.25 * log10(1 + c*x^2)$

and the magnitude loss due to trailing is given by

$DM_{SNR} = 1.25 * log10((1 + a*x^2)/(1 + b*x))$,

where the trail length, *x, is given by* $x = v * t_{exp} / Q$, in units of the seeing disk, *Q*, with *v* being the rate of motion and $t_{exp}$ the exposure time.

For LSST, the parameters *a = 0.76, b=1.16* and *c=0.42* have been determined (Jones et al., 2015). For the Flyeye telescope these parameters will need to be empirically determined. However, for the purpose of this investigation, the same parameters as for LSST were used. Assuming LSST-like losses may lead to the estimated trailing loss to be off by around 10%, which is relatively small and would apply equally to all Flyeye telescopes, thus not affecting the outcome of the site selection investigation.

Additionally, significant trailing is typically only observed in the last few days before impact, when the object is simultaneously brightening sharply. Therefore, the number of occasions where an object would drop below a detectability threshold due to the magnitude loss from trailing is low.

*2.3.3 Limitations*

For pointings obtained from the MPC, only the individual pointing positions were available, whereas the number of pointings and the time the observations were made are not. Astronomical surveys normally perform 4 observations of an individual pointing and require 4 observations of an asteroid to count as a single successful observation. We assumed that within the timescale of a night, the asteroid does not move fast enough to exit the field of view between pointings, and therefore assumed that if an asteroid was observed in one of the MPC pointings, it was observed at least 4 times to count as an observation. However, this may not necessarily be the case, for example if the observatory had to close halfway through the night due to bad weather, therefore, depending on the strategy employed by the individual survey, the number of detections from the MPC observatories may be overestimated.

Pointings from the MPC correspond to actual past pointings and contain the effect of the weather and closures due to technical maintenance. Simulated pointings from LSST have closures due to weather and maintenance included, however, the simulated pointings for the Flyeye do not include weather. For the Flyeye, we assumed a 10% loss of observations due to technical maintenance, though this percentage may increase depending on the actual maintenance requirements of the telescope. Therefore, the discovery rate for the Flyeye telescope may be overestimated.

## 3. Results and Discussion

### *3.1 Existing Telescopes*

Table 3 shows the summary of the results of the simulations for the 6 currently existing survey telescopes, along with their properties for ease of comparison. The uncertainties on the counts are given by √N, where N is the number of measurements, and the uncertainties on the timing are found from σ/√N, where σ is the standard deviation.

|  | Catalina 703 | Catalina G96 | PS-1 F51 | PS-2 F52 | ATLAS T05 (Haleakala) | ATLAS T08 (Mauna Loa) | ATLAS M22 (Sutherland) | ATLAS W68 (El Sauce) |
|---|---|---|---|---|---|---|---|---|
| **FoV diameter (deg)** | 4.4 | 2.2 | 3.0 | 3.0 | 5.4 | 5.4 | 5.4 | 5.4 |
| **Magnitude limit in *V*** | 19.7 | 21.5 | 22.5 | 22.5 | 19.6 | 19.6 | 19.6 | 19.6 |
| **Exposure time (s)** | 30 | 30 | 45* | 45* | 30 | 30 | 30 | 30 |
| **Seeing (")** | 5 | 3 | 1.2 *r* | 1.2 *r* | 4 | 4 | 4.9 | 4.3 |
| **No. Impactors discovered before impact H=25** | 494 ± 22 | 433 ± 21 | 377 ± 19 | 338 ± 18 | 560 ± 24 | 615 ± 25 | 438 ± 21 | 611 ± 25 |
| **Avg. days before impact H=25** | 5.4 ± 0.2 | 12.2 ± 0.5 | 14.8 ± 0.7 | 13.1 ± 0.6 | 3.9 ± 0.2 | 3.8 ± 0.1 | 5.1 ± 0.3 | 5.2 ± 0.2 |
| **No. Impactors discovered before impact H=28** | 62 ± 8 | 117 ± 11 | 88 ± 9 | 97 ± 10 | 48 ± 7 | 44 ± 7 | 31 ± 6 | 58 ± 8 |
| **Avg. days before impact H=28** | 2.5 ± 0.2 | 3.3 ± 0.3 | 4.8 ± 0.6 | 4.2 ± 0.4 | 2.2 ± 0.2 | 1.9 ± 0.2 | 2.4 ± 0.2 | 2.6 ± 0.3 |

Table 3. Number of impactors that would be detected by each of the existing surveys.
*Pan-STARRS produces images of exposure times between 30 s – 60 s; for these simulations we took every image as having an exposure time of 45 s, representing the average of this range.

Figures 3 and 4 show this same information as histograms for comparison.

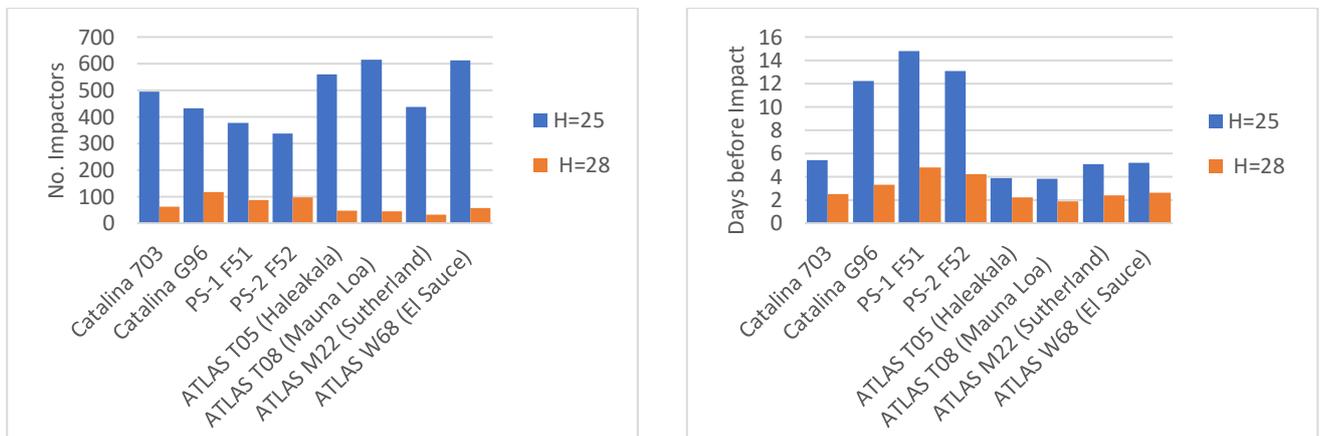

Figure 3. Number of impactors predicted to be discovered before impact by current telescopes for objects with H=25 and H=28.

Figure 4. Simulated average discovery time for current telescopes in days before impact for objects with H=25 and H=28.

For the larger, H=25 objects, the results show that currently existing telescopes have roughly similar discovery rates, which scale with the field of view of the respective survey telescopes. The ATLAS telescopes have the largest FoV, and correspondingly highest detection rates. M22 performs slightly worse than the other ATLAS units, but this could be because it is the first year of its operation. The two Pan-STARRS telescopes have an identical performance to each other within error but are slightly worse when compared to Catalina. When it comes to warning time in days, the fainter the limiting magnitude the better. For the smaller, H=28 objects, ATLAS does considerably worse than the others. This is because the smaller impacting objects are much fainter and will only get brighter than the magnitude limit in the last few days before impact, when there may not be enough time to detect them.

## 3.2 Flyeye & LSST

The same procedure was carried out for Flyeye and LSST as for the currently existing telescopes. Table 4 shows the summary of the simulations.

|  | LSST | Flyeye Mufara | Flyeye 809 | Flyeye B31 | Flyeye Muf ×2 | Flyeye Muf+809 | Flyeye Muf+B31 |
|---|---|---|---|---|---|---|---|
| **FoV diameter (deg)** | 3.5 | 6.7* | 6.7* | 6.7* | 6.7* | 6.7* | 6.7* |
| **Magnitude limit in *V*** | 25.4 | 21.5 | 21.5 | 21.5 | 21.5 | 21.5 | 21.5 |
| **Exposure time (s)** | 15 | 60 | 60 | 60 | 60 | 60 | 60 |
| **Seeing (")** | 0.7 | 1.5 | 1.5 | 1.5 | 1.5 | 1.5 | 1.5 |
| **No. Impactors discovered before impact H=25** | 760 ± 28 | 837 ± 29 | 822 ± 29 | 876 ± 30 | 962 ± 31 | 1167 ± 34 | 1159 ± 34 |
| **Avg. days before impact H=25** | 24.4 ± 0.6 | 12.1 ± 0.3 | 11.5 ± 0.4 | 12.0 ± 0.3 | 12.4 ± 0.3 | 12.0 ± 0.3 | 11.8 ± 0.3 |
| **No. Impactors discovered before impact H=28** | 314 ± 18 | 302 ± 17 | 255 ± 16 | 290 ± 17 | 461 ± 21 | 430 ± 21 | 436 ± 21 |
| **Avg. days before impact H=28** | 8.3 ± 0.5 | 3.8 ± 0.2 | 3.2 ± 0.2 | 3.9 ± 0.2 | 3.7 ± 0.2 | 3.5 ± 0.2 | 3.9 ± 0.2 |

Table 4. Number of impactors that would be detected by LSST and Flyeye in different locations.

Figures 5 and 6 show this same information as histograms for comparison.

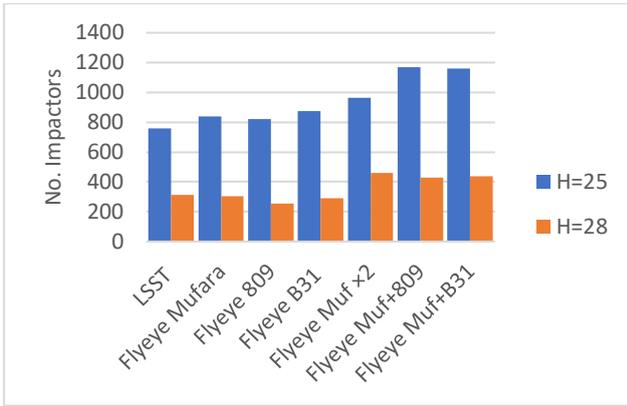
Figure 5. Number of impactors predicted to be discovered before impact by future telescopes for objects with H=25 and H=28.

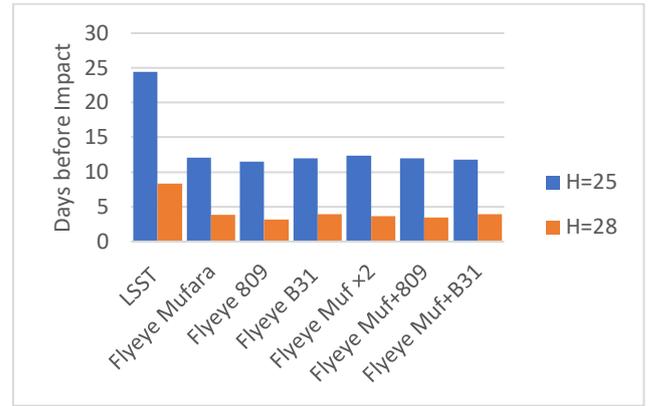
Figure 6. Simulated average discovery time for future telescopes in days before impact for objects with H=25 and H=28.

For each detection, we assume counting statistics with an uncertainty of $\sqrt{N}$. When comparing the number of detections of the Flyeye at the Mufara, La Silla, and Sutherland sites, we find that the differences are effectively zero within an uncertainty of approximately one standard deviation. This is true for both the larger and smaller objects. Having two Flyeye telescopes in the same location, shown by "Mufara ×2", increases the number of discoveries by around 50%, which is statistically significant but not a factor of two, due to reaching completness of coverage. While the increase in detections is not proportional to the number of telescopes, there is some benefit of improving the cadence of observations.

The above analysis only considers detections by each site in isolation; the uniqueness of these detections is examined in the following section.

For both larger and smaller objects, LSST finds a comparable number of objects to the Flyeye, but many more days in advance. The relative low number of discoveries in comparison to the expected limiting magnitude is likely due to the irregular observing strategy employed by LSST. This will be still further optimised as the telescope comes online.

*3.3 The Impact of LSST on Flyeye Telescope Detections*

In this section we present the results on whether the operation of LSST would have an impact on the detection rate of the Flyeye. Two scenarios were investigated: one, before the LSST is operational, and one after. The detection rate of the single Flyeye telescopes and combinations of 2 Flyeye telescopes were compared to the rates of existing telescopes in both scenarios. Of particular interest was the number of unique objects found by each telescope, i.e. a detection that none of the other telescopes made, labelled "Unique detections" in Tables 5 – 6. Another parameter that was investigated was the number of "first detections", i.e. the number of times a given telescope is the first to make a detection when the detection is not unique. The results of this are summarised in Table 5. for H=25 and Table 6. for H=28.

In the first 4 rows we present the combined results from existing NEO survey telescopes with the upcoming ones in different scenarios: LSST alone, Flyeye-1 alone, Flyeye-2 alone in the location under analysis, and Flyeye-1 & -2 operational.

In the last 3 rows, we assume that LSST is fully operational and contributing to the of the overall network of existing telescopes, which is then compared to adding 1 or 2 Flyeye telescopes.

|  | LSST | Flyeye Muf | Flyeye 809 | Flyeye B31 | Flyeye Muf ×2 | Flyeye Muf+809 | Flyeye Muf+B31 |
|---|---|---|---|---|---|---|---|
| Total objects detected with all telescopes | 1354 ± 37 | 1406 ± 37 | 1470 ± 38 | 1470 ± 38 | 1471 ± 38 | 1577 ± 40 | 1580 ± 40 |
| Total detections by this telescope | 760 ± 28 | 837 ± 29 | 822 ± 29 | 876 ± 30 | 962 ± 31 | 1159 ± 34 | 1167 ± 34 |
| First detections | 554 ± 24 | 433 ± 21 | 388 ± 20 | 421 ± 21 | 512 ± 23 | 553 ± 24 | 559 ± 24 |
| Unique detections | 82 ± 9 | 134 ± 12 | 198 ± 14 | 198 ± 14 | 199 ± 14 | 305 ± 17 | 308 ± 18 |
| After LSST: Total by all | -- | 1476 ± 38 | 1504 ± 39 | 1496 ± 39 | 1539 ± 39 | 1609 ± 40 | 1604 ± 40 |
| After LSST: First | -- | 264 ± 16 | 180 ± 13 | 186 ± 18 | 321 ± 18 | 330 ± 18 | 333 ± 18 |
| After LSST Unique | -- | 122 ± 11 | 150 ± 12 | 142 ± 14 | 185 ± 14 | 255 ± 16 | 250 ± 16 |

Table 5. Summary of detections from LSST and Flyeye telescope combinations for objects with H=25.

|  | LSST | Flyeye Muf | Flyeye 809 | Flyeye B31 | Flyeye Muf ×2 | Flyeye Muf+809 | Flyeye Muf+B31 |
|---|---|---|---|---|---|---|---|
| Total objects detected with all telescopes | 517 ± 23 | 491 ± 22 | 464 ± 22 | 488 ± 22 | 590 ± 24 | 573 ± 24 | 583 ± 24 |
| Total detections by this telescope | 314 ± 18 | 302 ± 17 | 255 ± 16 | 290 ± 17 | 461 ± 21 | 430 ± 21 | 436 ± 21 |
| First detections | 103 ± 10 | 71 ± 8 | 53 ± 7 | 67 ± 8 | 107 ± 10 | 89 ± 9 | 90 ± 9 |
| Unique detections | 181 ± 13 | 155 ± 12 | 128 ± 11 | 152 ± 12 | 254 ± 16 | 237 ± 15 | 247 ± 16 |
| After LSST: Total by all | -- | 632 ± 25 | 604 ± 25 | 614 ± 26 | 703 ± 27 | 693 ± 26 | 695 ± 26 |
| After LSST: First | -- | 47 ± 7 | 31 ± 6 | 42 ± 6 | 79 ± 9 | 60 ± 8 | 60 ± 8 |
| After LSST Unique | -- | 115 ± 11 | 87 ± 9 | 97 ± 10 | 186 ± 14 | 176 ± 13 | 178 ± 13 |

Table 6. Summary of detections from LSST and Flyeye telescope combinations for objects with H=28.

In the case of the larger, H=25, objects, the total detection rate for all telescopes is of around 1400 for an individual Flyeye telescope in each location, with no statistically significant difference between the north and the south. After LSST, the total number of detections will increase by the number of unique detections from LSST.

Figure 7 represents the same information as in the two tables in a histogram format for visual comparison.

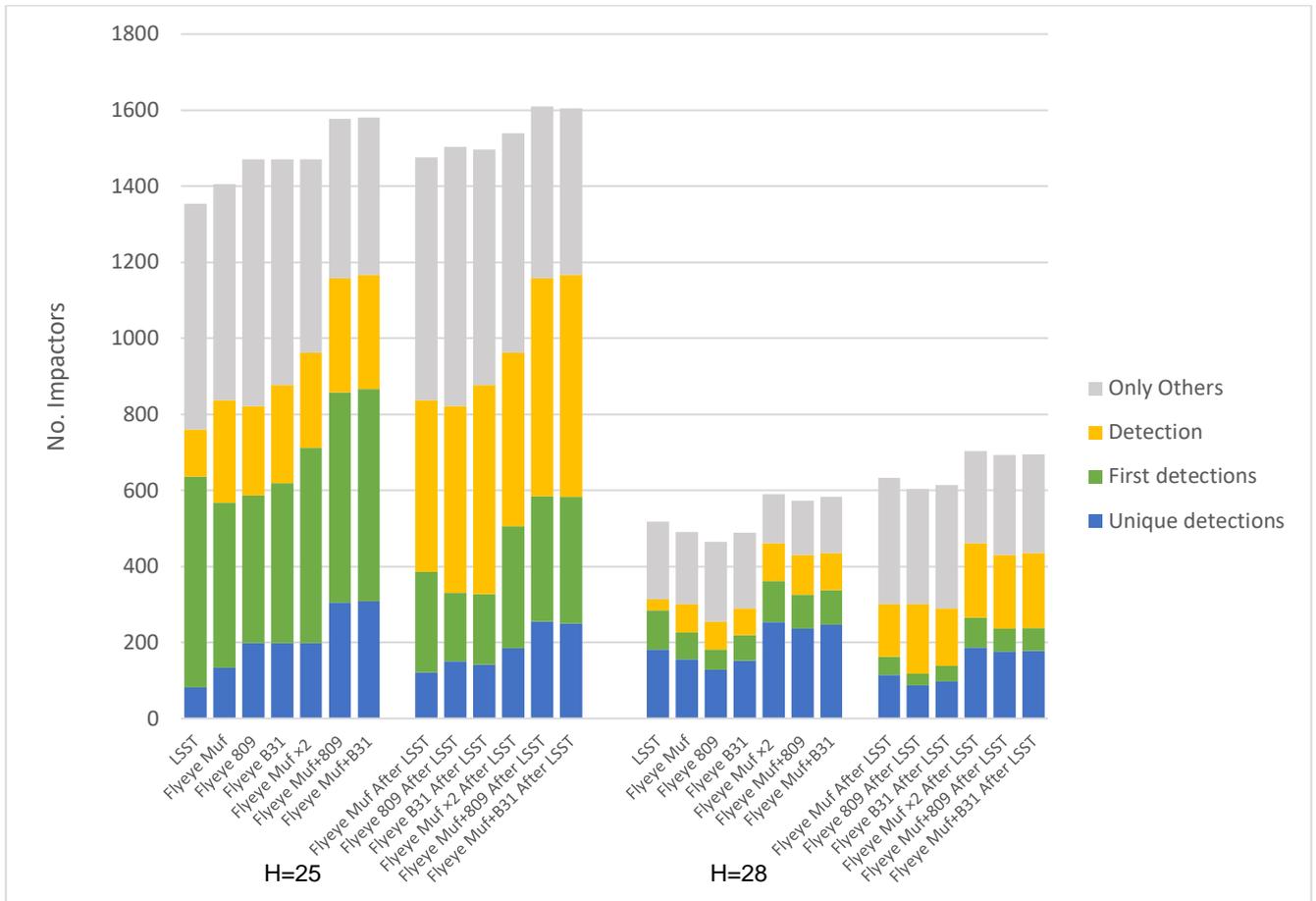

Figure 7. Histogram showing breakdown of details of impactors detected by future telescopes before and after LSST for objects with H=25 and H=28.

The results show that, before LSST, for the larger objects, a single Flyeye will discover more unique objects if it is placed in the south. This is due to the large number of surveys operating in the north. The number of "first detections" is roughly the same for all sites. In comparison to the currently existing surveys, a single Flyeye installed on Mount Mufara in Italy would increase the detection rate by 11% for larger asteroids and 46% for smaller asteroids in the absence of LSST. For the objects detected by the Flyeye, they will be found on average 12 days before impact for the larger objects and 4 days before impact for the smaller ones. This is a much earlier warning time than ATLAS, which was found to give on average a warning time of 5 and 2 days for these objects respectively. The simulations also show that, for the larger objects, having one Flyeye in the north and a second Flyeye in the south will lead to 24% more discoveries compared to the current baseline, while having two Flyeye telescopes in the north will only lead to 16% more discoveries. After LSST, the unique number of discoveries from the Flyeye will drop, particularly in the south. However, it will only result in the Flyeye telescopes in north and south to have the same discovery rate within around a one-sigma uncertainty. When considering a combination of two Flyeye telescopes in the same location, doubling the cadence will increase the number of unique detections by around 50%, while having a combination of one Flyeye in the north and one in the south will more than double the discovery rate, due to an increased sky coverage. (As seen in Table 5 and Figure 7, the numbers are 134 ± 12 for a single Flyeye in the north, 199 ± 14 for two Flyeye telescopes in the north, 305 ± 17 and 308 ± 18 for a Flyeye in the north plus 809 and B31, respectively.) Even after LSST, the number of unique impactor detections from having one telescope in the Northern Hemisphere and one in the Southern Hemisphere will still be greater than from two telescopes in the Northern Hemisphere. (After LSST the number of unique detections for two Flyeyes is 185 ± 14, still around 50% more compared to a single Flyeye, while it is 255 ± 16 and 250 ± 16 for north plus 809 and B31, respectively, still around double compared to the number of unique detections from a single telescope, 122 ± 11.) When combined with a telescope in the North, both La Silla (809) and Sutherland (B31) locations perform identically, within error. In this case, the number of clear nights will be a determining factor when choosing between sites.

An interesting observation is that for H=25 objects, LSST rarely appears to be "unique". Most likely this is because of its observing cadence, it is missing some of the fastest objects, that the others may get. Also note that since about half of the impactors will be impacting from the Sun-direction, they will not be able to be detected by any ground-based telescope. Therefore, at around 1600 impactors the observations are mostly complete at this size.

The fainter, H=28 objects are much more difficult to observe, and there are fewer detections of these objects before impact. The likelihood that the detection is unique to a telescope is greater. For these smaller objects, the positioning between north and south makes no difference, within error. This indicates that most surveys are currently not reaching completeness at the fainter end, i.e. they are not detecting a large number of the smaller impacting asteroids. Since LSST will also observe some of the same asteroids as the Flyeye, it will reduce the Flyeye's number of unique asteroid discoveries. However, there is no statistically significant difference in the change between the Northern and the Southern Hemispheres. For these objects, there is also no difference in terms of unique discoveries between having two co-located Flyeye telescopes or one in each hemisphere.

## 4. Weather Effects

The main limitation of the simulations is that it does not take weather into account. A brief summary of the weather is presented below.

According to the European Southern Observatory (ESO) (e.g. European Southern Observatory, 2013) and other institutions owning telescopes in the La Silla Observatory (e.g. Jehin et al., 2014), the site experiences about 300 clear nights per year. A quick calculation from their publicly available historical monthly weather reports between 1991-1999[4], an average of 15.9% of nights are categorised as "useless", corresponding to 307 astronomically useful nights, which confirms their claim.

According to the South African Astronomical Observatory (SAAO) and telescopes operating there (e.g., report by Booth et al., 2020), the site experiences around 250 clear nights per year on average.

Astronomical weather data for Mt. Mufara are harder to find. Weather sites for Sicily[5] give the average number of clear and mostly clear skies as 68%, which would correspond to 248 nights per year. Analysis of satellite data (Cavazzani et al., 2011) showed 72.5% of clear nights at La Palma. We can assume Sicily will have a similar or slightly worse percentage. The same study showed 76% of satellite clear nights for La Silla.

Including weather losses for the Flyeye telescopes was beyond the scope of this investigation, as it would have introduced additional complexities and potential biases if handled incorrectly. However, it is important to note that La Silla experiences significantly more clear nights compared to the other sites. Therefore, if the simulations show that a similar or greater number of impactors can be detected from La Silla without considering the weather, even in the presence of LSST, then including the weather effects would only further improve La Silla's position as the top site.

## 5. Conclusions

The results of the simulations show no clear statistical evidence in favour of North vs South for a single Flyeye telescope. When considering two Flyeye telescopes, the simulations do show that, for larger objects, having two complementary telescopes, one in the north and one in the south, will make more discoveries than two telescopes in the same location in the north. However, no longitudinal difference was found between the southern sites. For the smaller objects, there is no statistical difference between sites. The simulations also show that while LSST will be the first to discover many asteroids, particularly in the Southern Hemisphere, it will not overwhelm the number of asteroid detections made by other telescopes, including the Flyeye.

---

[4] https://www.eso.org/sci/facilities/lasilla/astclim/weather/tablemwr.html
[5] https://weatherspark.com/y/150410/Average-Weather-in-Sicily-Italy-Year-Round

Combining the results of the simulations with the analysis of the weather, which indicates that La Silla has the highest number of clear nights per year, we conclude that La Silla would be the best site for Flyeye-2 among those analysed.